\documentclass[aip,apl,amsmath,amssymb,amsfonts,reprint,a4paper,superscriptaddress]{revtex4-1}
\usepackage{newtxtext,newtxmath}
\usepackage[T1]{fontenc}
\usepackage[utf8]{inputenx}
\usepackage{graphicx}
\usepackage[dvipsnames]{xcolor}
\usepackage{soul} 
\usepackage[textwidth=17.5cm,textheight=23.5cm,verbose,pdftex]{geometry}
\usepackage[pdftex]{hyperref}
\usepackage{booktabs} 

\hypersetup{pdfauthor={Johannes Feldl}, bookmarksnumbered=true, pdftitle={Magnetic characteristics of epitaxial NiO films studied by Raman spectroscopy}, colorlinks, citecolor=blue, linkcolor=blue, urlcolor=blue}

\begin{document}
\title{Magnetic characteristics of epitaxial NiO films studied by Raman spectroscopy}

\author{J. Feldl} 

\email[Electronic mail: ]{feldl@pdi-berlin.de}

\author{M. Budde}
\author{C. Tschammer}
\author{O. Bierwagen}
\author{M. Ramsteiner}
\email[Electronic mail: ]{ramsteiner@pdi-berlin.de}
\affiliation{Paul-Drude-Institut für Festkörperelektronik, Leibniz-Institut im Forschungsverbund Berlin e.\,V., Hausvogteiplatz 5--7, 10117 Berlin, Germany}

\begin{abstract}
Raman spectroscopy is utilized to study the magnetic characteristics of heteroepitaxial NiO thin films grown by plasma-assisted molecular beam epitaxy on MgO(100) substrates. For the determination of the N{\'e}el temperature, we demonstrate a reliable approach by analyzing the temperature dependence of the Raman peak originating from second-order scattering by magnons. The antiferromagnetic coupling strength is found to be strongly influenced by the growth conditions. The low-temperature magnon frequency and the N{\'e}el temperature are demonstrated to depend on the biaxial lattice strain and the degree of structural disorder which is dominated by point defects.
\end{abstract}

\maketitle

\section{\label{sec:level1}Introduction\protect\\}

\noindent 

NiO is one of the rare transparent oxides (bandgap around
3.7~eV) which can be utilized as a \textit{p}-type semiconductor.\citep{ohta2003a,rao1965a,zhang2018a} Furthermore, as one of the most common antiferromagnetic oxides (with an antiferromagnetic to paramagnetic phase-transition at a N{\'e}el temperature of 523~K) NiO is of interest for both fundamental studies and spintronic applications.\citep{pinarbasi2000a,moriyama2018a} NiO has a rock salt structure with ferromagnetic alignment of the magnetic moments in each $\{$111$\}$ plane and antiferromagnetic coupling between neighboring $\{$111$\}$ planes via a Ni-O super exchange interaction.\citep{hutchings1972a} Thereby, the next-nearest-neighbor (180$^\circ$) Ni$^{+2}-$O$^{-2}-$Ni$^{+2}$ antiferromagnetic exchange $J$ is much larger than the nearest-neighbor (90$^\circ$) ferromagnetic exchange making the latter one negligible.\citep{massey1990a} However, in thin magnetic films  synthesized by various methods, the ferromagnetic or antiferromagnetic coupling is expected to be modified and deteriorated by disorder and strain as reported previously for ferromagnetic manganites.\citep{doerr2006a}  Regarding lattice strain, a well-defined influence on the magnetic characteristics is expected based on results obtained by measurements under hydrostatic pressure.\citep{massey1990a} For the experimental investigation of structural and magnetic properties, Raman spectroscopy of phonons and magnons in NiO \citep{dietz1971a,dietz1971b,massey1990a,lockwood1992a,pressl1996a,grimsditch1998a,mita2002a,
mironova2007a,gandhi2011a,mironova2011a,duan2012a,marcius2012a,gandhi2013a,aytan2017a,lacerda2017a,
budde2018a} and other antiferromagnetic materials \citep{fleury1968a,fleury1970a,dietz1971a,lockwood1975a,schilbe1997a} has been proven to be a powerful method. In addition to the Raman spectroscopic work, theoretical studies \citep{cottam1972a,reichardt1975a,cottam1979a,kushwaha1982a,floris2011a} on phonons and magnons in NiO as well as investigations based on other experimental techniques \citep{hutchings1972a,
grimsditch1994a,uchiyama2010a,willett2015a} can be found in the literature.

In this work, Raman spectroscopy is demonstrated as a tool to investigate both the structural and magnetic characteristics of NiO thin films. As a result, we reveal that the low-temperature magnon frequency and the N{\'e}el temperature depend not only on the biaxial misfit strain but also on structural disorder in the epitaxial NiO films.
\begin{figure}[h]
\includegraphics*[width=0.48\textwidth]{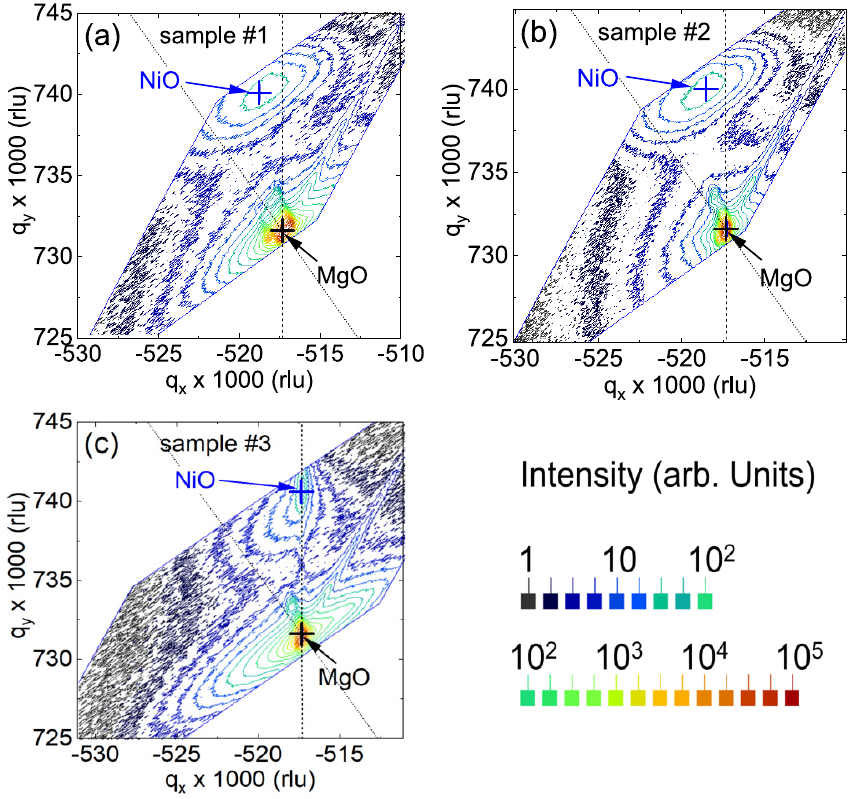}
\caption{X-ray diffraction reciprocal space maps around the 422 reflection of (a) sample \#1 grown at 700~$^\circ$C, (b) sample \#2 grown at 200~$^\circ$C and (c) sample \#3 grown at 20~$^\circ$C. The maps are drawn by equiintensity lines using a color coded, logarithmic scale as specified. The numerically extracted peak positions of the MgO substrate and the thin NiO films are marked by black and blue crosses, respectively, and labeled accordingly. The unit rlu is defined as $\lambda/2d$ with x-ray wavelength $\lambda$ and spacing of the lattice planes $d$. $q_x$ and $q_y$ correspond to the in-plane (022) and out-of plane (400) components of the investigated reflection. A NiO peak on the vertical dashed or oblique dotted line corresponds to pseudomorphically strained or fully relaxed layers, respectively.} 
\label{RSM}
\end{figure}
\section{EXPERIMENTAL DETAILS}
\begin{figure}[h]
\includegraphics*[width=0.45\textwidth]{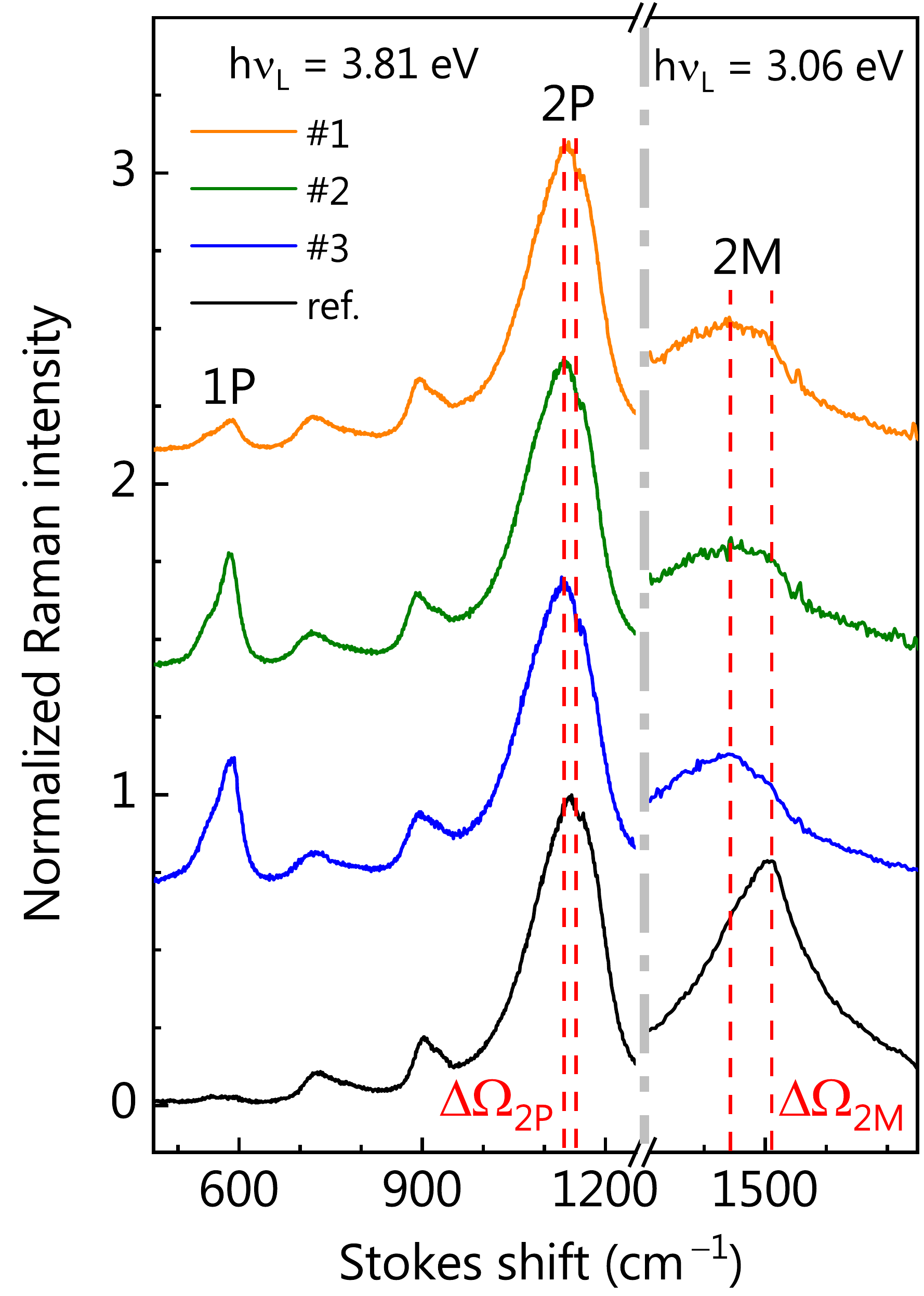}
\caption{Room temperature Raman spectra of the thin NiO films
and a reference bulk NiO sample excited at 3.81~eV (lower frequency range)
and 3.06~eV (frequency range above 1220~cm$^{-1}$).} 
\label{Spectra}
\end{figure}

The investigated NiO samples were grown on MgO(100) substrates by plasma-assisted molecular beam epitaxy. Because of the common rock-salt crystal structure and the similar lattice constants of MgO (0.4212~nm) and NiO (0.4176~nm), MgO is a widely used, suitable substrate for the growth of high quality NiO films.\cite{zhang2018a} Covering a large range of substrate temperatures, the investigated films were grown at 700~$^\circ$C (sample~\#1), 200~$^\circ$C (sample~\#2) and 20~$^\circ$C (sample~\#3). These samples correspond to S2-700, S2-200 and S2-20 of Ref.~\onlinecite{budde2018a}. Films grown at higher substrate temperatures exhibit Ni$_x$Mg$_{1-x}$O alloying. 
From the chosen growth conditions and the previous characterization of the NiO films, the existence of secondary phases and a significant impact of surface roughness can be excluded for all samples (see Ref. \onlinecite{budde2018a}). The investigated single crystalline thin NiO films have thicknesses of about 60~nm, a close-to-stoichiometric composition and exhibit optical transparency in the visible spectral range. For reference purposes, an about 0.7~mm thick green NiO foil has been utilized as a bulk-like sample. This sample has been characterized previously by x-ray diffraction, spectroscopic ellipsometry and Raman spectroscopy (see Ref.~\onlinecite{budde2018a}). The results reveal that our reference sample can be regarded as unstrained bulk NiO with state-of-the-art quality. The occurence of a weak Raman-forbidden peak at 580~cm$^{-1}$ (see Fig.~\ref{Spectra}) is commonly observed for single crystalline NiO.\citep{aytan2017a,mironova2007a,mironova2011a} 

The in-plane and out-of-plane lattice parameters of the thin NiO films were determined by x-ray diffraction reciprocal space mapping (RSM). The corresponding
 RSMs were taken around the 422 reflection in asymmetric geometry by a four-cicle lab-diffractometer (PANalytical X$'$Pert Pro MRD) using a line detector (Pixcel, 256 channels covering a 2$\Theta$-range of 2.51$^\circ$) and Cu K$\alpha$ radiation.

The Raman spectroscopic measurements were performed in the backscattering configuration with optical excitation at wavelengths of 405~nm (solid-state laser) and 325~nm (He-Cd laser). The incident laser light was focused by a microscope objective onto the sample surface (spot diameter of approximately 2~$\mu$m). The backscattered light was collected by the same objective (confocal configuration), spectrally dispersed by an 80-cm spectrograph (LabRam HR Evolution, Horiba/Jobin   Yvon) and detected with a liquid-nitrogen-cooled charge-coupled device (CCD). For Rayleigh light suppression, a bandpass filter with ultra-narrow spectral bandwidth was used. For the temperature dependent Raman scattering measurements, a continous-flow cryostat (CryoVac) and a heating stage (Linkam) were used in the temperature ranges of $10$--$300$~K and $300$--$550$~K, respectively.

\section{RESULTS and DISCUSSION}

\subsection{\label{sec:level2}Lattice strain}
Figure~\ref{RSM} shows the RSMs of the thin NiO films. The in-plane ($a$) and out-of-plane ($c$) lattice parameters of the NiO films, extracted from these maps using the relation $q = \lambda /2d$ with wavelength $\lambda$ and spacing of the crystal planes $\left( d_{400}=c/4, d_{022}=a/\sqrt(8) \right)$, are shown in Table \ref{tab:samples}. Interestingly, growth at 20~$^\circ$C resulted in a pseudomorphically strained NiO layer, indicated by the coinciding in-plane lattice parameter (same $q_{\text{x}}$) of NiO film and MgO substrate in Fig.~\ref{RSM}(c). Higher growth temperatures of 200~$^\circ$C and 700~$^\circ$C, in contrast, resulted in the growth of thin NiO films that are partially relaxed (towards the smaller lattice parameter of NiO than that of MgO), indicated by larger magnitude $q_{\text{x}}$ of the layer compared to that of the substrate in Fig.~\ref{RSM}(b,a).

\begin{table}
\caption{\label{tab:samples}Substrate growth temperature $T_{\text{S}}$, in- and out-of-plane lattice parameter $a$ and $c$ as well as the quality index 1$-$Q for the bulk reference and samples \#1 to \#3.}
\begin{ruledtabular}
\begin{tabular}{ccccc}
Sample&$T_{\text{S}}$ ($^\circ$C)& $a$ (\AA)& $c$ (\AA)&$1-Q$\\
\hline
\ ref. & N/A & 4.178 & 4.178 & 0.97\\
\#1 & 700& 4.200 & 4.163 & 0.88\\
\#2 & 200& 4.201 & 4.164 & 0.81 \\
\#3 & 20 & 4.211 & 4.160 & 0.77 \\
\end{tabular}
\end{ruledtabular}
\end{table}

\subsection{\label{sec:level2}Structural disorder}
Figure~\ref{Spectra} displays room temperature Raman spectra of the thin NiO films and the reference bulk NiO sample. Optical excitation at 3.81~eV (325~nm) close to the NiO bandgap was used for the lower frequency range (up to 1220~cm$^{-1}$). This condition enables a high sensitivity for optical-phonon scattering due to resonance-enhanced Raman scattering. The corresponding spectrum of bulk NiO exhibits only peaks due to second-order phonon scattering at 730, 900, and 1130~cm$^{-1}$ (2P) in agreement with the forbidden first-order Raman scattering by optical phonons in the rock-salt structure.\citep{dietz1971a, aytan2017a} In contrast, the thin NiO films reveal also first-order Raman scattering at 580~cm$^{-1}$ (1P).\citep{mironova2007a,aytan2017a} As a further deviation from the spectrum of bulk NiO, the 2P peak of the thin NiO films is shifted to lower frequencies. This redshift $\Delta\Omega_\text{2P}$ is attributed to biaxial strain in the thin NiO films.  

Since the occurence of the Raman peak 1P is enabled by a deviation from the perfect lattice order, we have defined a quality index $Q$ as an inverse figure of merit for the crystal quality:\citep{budde2018a} 
\begin{equation}
Q = \frac {I_{\mathrm{1P}}} {I_{\mathrm{2P}}}.
\label{eq:one}
\end{equation}
Here, $I_\text{1P}$ and $I_\text{2P}$ denote the intensity of the 1P and 2P Raman peaks, respectively. Note that $Q = 0$ corresponds to a perfect NiO crystal with rock-salt structure and $1-Q$ reflects the degree of structural order in NiO. As shown in Fig.~\ref{Q_and_2P_shift}, the magnitude of $1-Q$ decreases monotonically with the tensile in-plane strain determined from the data shown in Table \ref{tab:samples} assuming a lattice parameter of 4.176~\AA~ for bulk NiO. Since the pseudomorphically strained film (sample \#3) exhibits the lowest crystal quality $1-Q$, our result indicates that the lattice disorder in the thin NiO films is dominated by point defects rather than by misfit dislocations induced by partial relaxation of the misfit strain (in samples \#1 and \#2). In fact, oxygen rich growth conditions lead to the creation of acceptor-like Ni vacancies in NiO films.\citep{zhang2018a} Furthermore, low growth temperatures favor the formation of additional point defects or defect complexes as demonstrated in Ref.~\onlinecite{karsthof2019a} for NiO films synthesized by pulsed laser deposition. Consequently, it is reasonable to attribute the decreasing magnitude of $1-Q$ at low growth temperatures (see Table~\ref{tab:samples}) to an increasing density of point defects. The opposite strain-dependence of $1-Q$ can be observed for thin NiO films grown on c-plane GaN substrates,\citep{budde2020a} as shown in the inset of Fig.~\ref{Q_and_2P_shift}. This finding confirms that the quantity $1-Q$ indeed mainly reflects the crystal quality independent of the lattice distortion induced by the biaxial misfit strain.\citep{budde2018a,budde2020a}

\subsection{\label{sec:level2}Magnetic characteristics}
For the investigation of temperature-dependent second-order Raman scattering by magnons, non-resonant optical excitation at 3.06~eV is used in order to obtain a comparable signal strength for second-order phonon and magnon scattering. The corresponding room temperature Raman spectra shown in Fig.~\ref{Spectra} clearly reveal the presence of the second-order magnon peak (2M) at about 1450~cm$^{-1}$ for both, the bulk NiO sample and the thin NiO films.\citep{dietz1971a,massey1990a} However, the intensity of the 2M peak of the present thin NiO films is reduced (as verified by polarization dependent Raman\citep{mironova2011a} measurements) and its position is shifted by $\Delta\Omega_\text{2M}$ to lower frequencies. Both findings are explained by a reduction in the antiferromagnetic coupling via Ni-O super exchange interaction.\citep{mironova2007a} Note that for the above mentioned thin NiO films grown on GaN or nanosized NiO prepared by a plasma synthesis, no 2M peaks could be resolved in the Raman spectra at room temperature and, therefore, no signature of antiferromagnetism.\citep{mironova2007a, budde2020a}

\begin{figure}
\includegraphics*[width=0.46\textwidth]{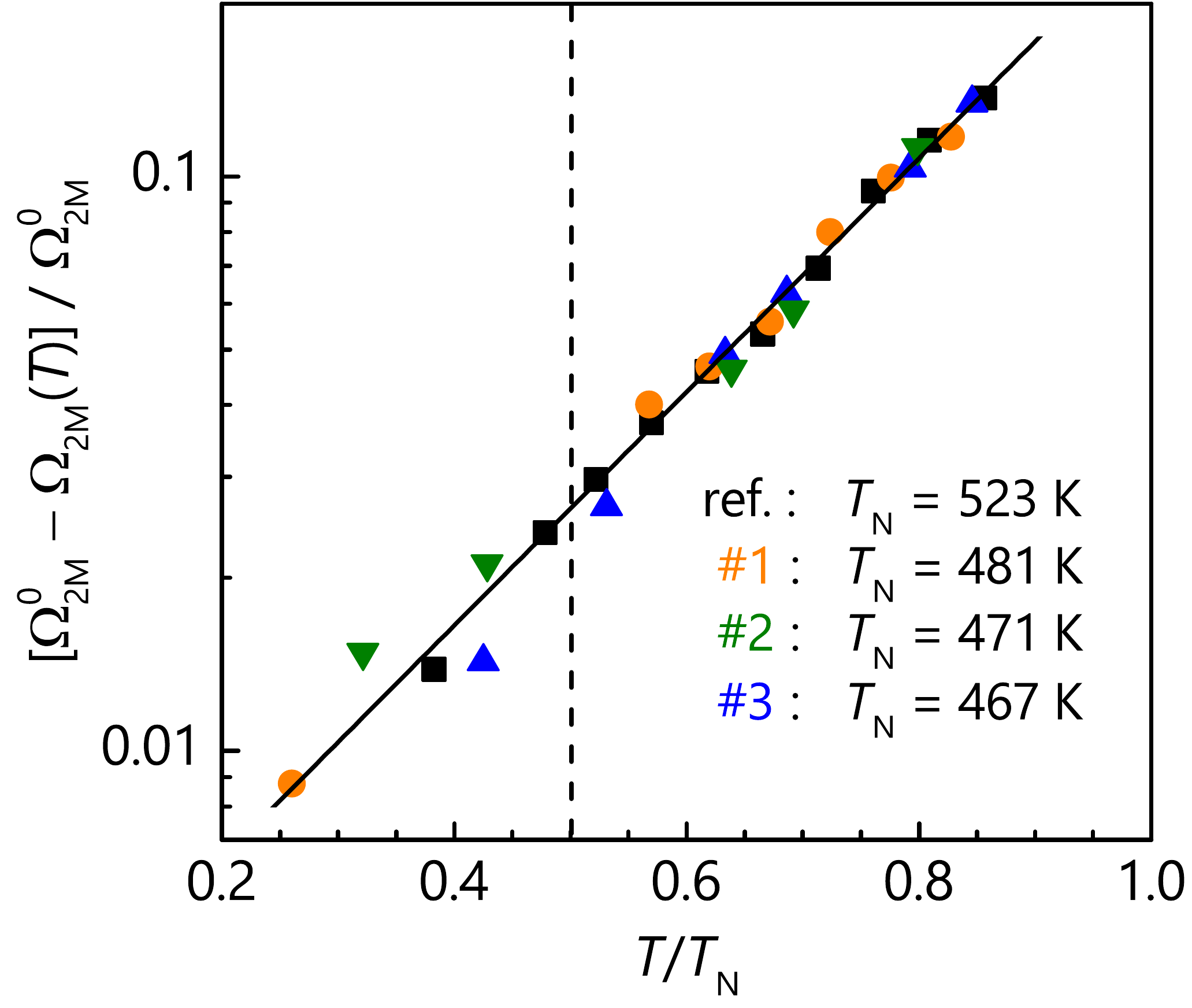}
\caption{Relative frequency shift of the 2M peak $[\Omega_\text{2M}^{0} - \Omega_\text{2M}(T)]/\Omega_\text{2M}^{0}$ for bulk NiO (ref.) and the
thin NiO films (samples \#1, \#2, \#3) as a function of the reduced
temperature $T/T_\text{N}$. The values of $T_\text{N}$ for the three thin NiO films
have been determined as described in the text. The solid line represents the
function $\theta \textrm{exp}(\beta T/T_\text{N})$ with $\theta = 2.54 \times$10$^{-3}$ and $\beta = 4.68$. The fit was performed in the range $0.5 \leq T/T_\text{N} \leq 1.0$ as indicated by the dashed line.}
\label{2M_relshift}
\end{figure}

An important characteristic of antiferromagnets is the N{\'e}el temperature ($T_\text{N}$). In order to determine $T_\text{N}$ for the investigated thin NiO films, we take advantage of the common relationship 
\begin{equation}
\frac{\Omega_\text{2M}^{0} - \Omega_\text{2M}(T)}{\Omega_\text{2M}^{0}} = \theta \exp\left( \beta \frac {T}{T_\text{N}}\right)
\label{eq:two}
\end{equation} 
between the relative shift of the temperature dependent 2M peak $[\Omega_\text{2M}^{0} - \Omega_\text{2M}(T)]/\Omega_\text{2M}^{0}$ and the reduced temperature $T/T_\text{N}$ with constant values of $\theta$ and $\beta$. $\Omega_\text{2M}^{0}$ is 
the respective low-temperature saturation value of the 2M peak frequency. This relationship has been demonstrated previously for different antiferromagnets (NiO and KNiF$_3$).\citep{dietz1971a} In  Fig.~\ref{2M_relshift} we show $[\Omega_\text{2M}^{0} - \Omega_\text{2M}(T)]/\Omega_\text{2M}^{0}$ as a function of $T/T_\text{N}$ for bulk NiO and the three thin NiO films. In the range $0.5 \leq T/T_\text{N} \leq 1.0$, all data points indeed follow the function given in Eq.~\ref{eq:two}. Within our approach, we first determine the parameters $\theta$ and $\beta$ by fitting the data of the reference bulk sample, keeping the well-known value of $T_\text{N} = 523$~K fixed (see Eq.~\ref{eq:two}). The obtained values $\theta$ and $\beta$ are then used for the fits of the data from individual thin NiO films with $T_\text{N}$ as the only free parameter. The excellent agreement between the data and the function given by Eq.~\ref{eq:two} demonstrates the consistency and reliability of our approach for the determination of the N{\'e}el temperature, an achievement which would otherwise be difficult to attain. The obtained values of $T_\text{N}$ for samples \#1, \#2 and \#3  are $481$~K, $471$~K and $467$~K, respectively. 

\begin{figure}[h]
\includegraphics*[width=0.43\textwidth]{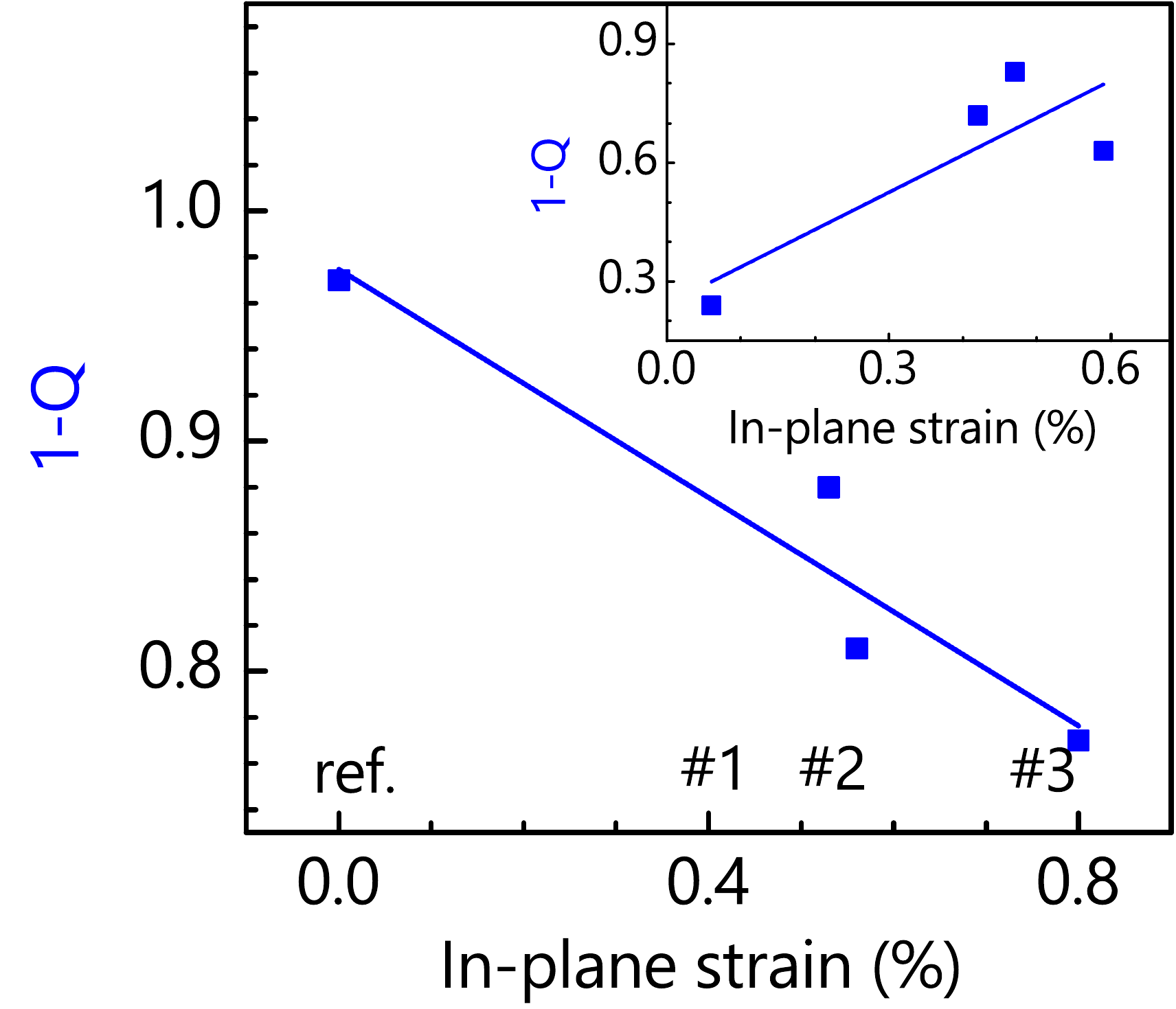}
\caption{Degree of structural order $1 - Q$ for
bulk NiO (ref.) and all three thin NiO films (samples \#1, \#2, \#3)
as a function of the in-plane strain. The inset shows $1 - Q$ for thin NiO films grown on GaN substrates. The straight lines are guides for the eye. } 
\label{Q_and_2P_shift}
\end{figure}

\begin{figure}
\includegraphics*[width=0.45\textwidth]{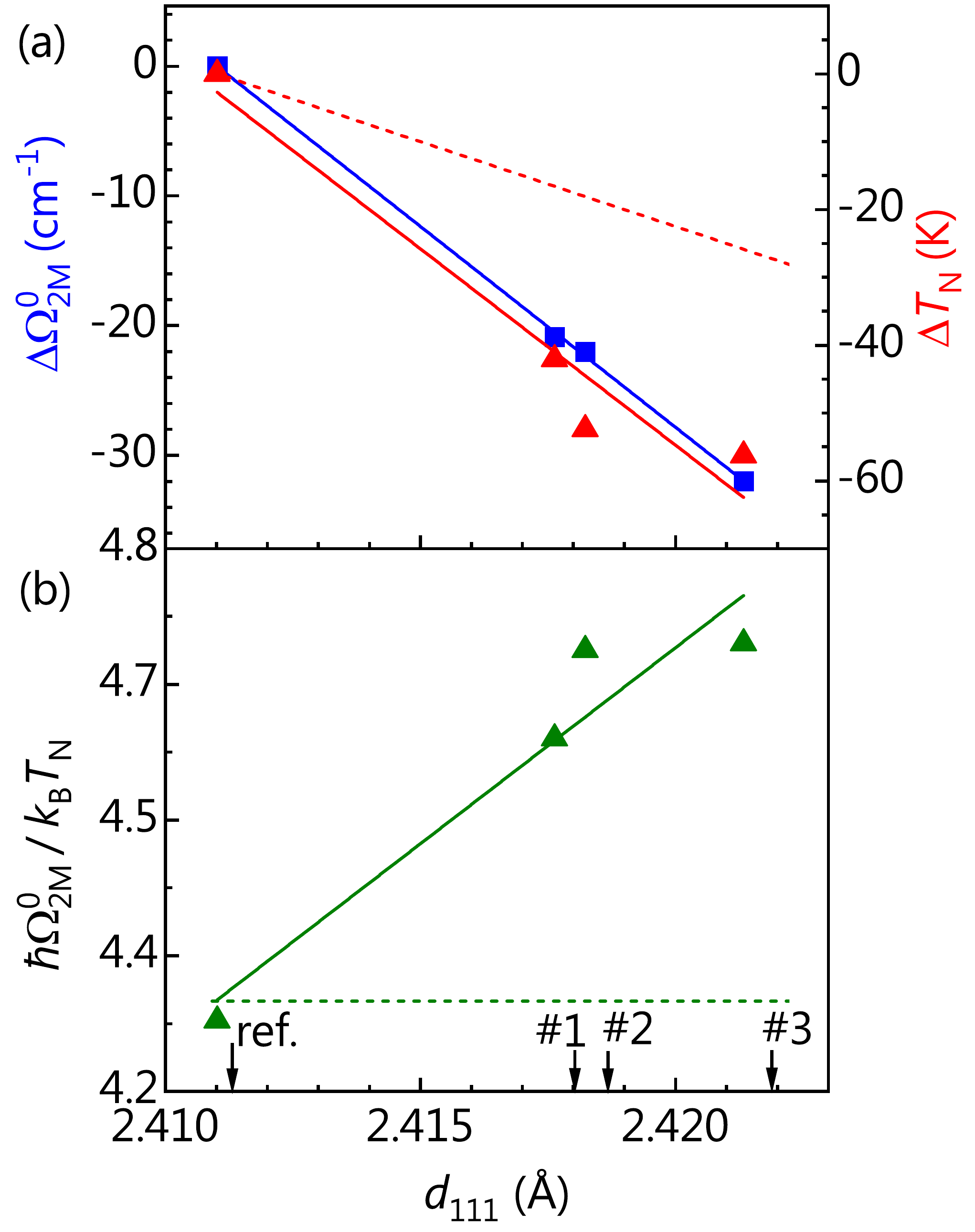}
\caption{(a) Shift of the low-temperature 2M peak frequency $\Delta\Omega
_\text{2M}^0$ as well as shift of the Néel temperature $\Delta T_\text{N}$ for bulk NiO (ref.) and all three thin NiO films (samples \#1, \#2, \#3) as a function of the $\{$111$\}$ plane spacing. The shift of the N{\'e}el temperature according to Ref.~\onlinecite{massey1990a} is shown for comparison (dashed line). (b) Ratio $\hslash\Omega_\text{2M}^0\,/\,k_{B}T_\text{N}$ for bulk NiO (ref.) and all three thin NiO films (samples \#1, \#2, \#3) as a function of the
$\{$111$\}$ plane spacing. The constant ratio according to Ref.~\onlinecite{massey1990a} is shown for comparison (dashed line).} 
\label{TN_and_2M_shift} 
\end{figure}

Fig.~\ref{TN_and_2M_shift}(a) displays the shift of the low-temperature 2M peak frequency $\left(\Delta\Omega_\text{2M}^{0} = \Omega_\text{2M}^{0,\text{film}} - \Omega_\text{2M}^{0,\text{bulk}}\right)$ as well as the shift of the Néel temperature ($\Delta T_\text{N}$) as a function of the separation $d_{111}$ between $\{$111$\}$ lattice planes which is the crucial distance regarding the strength of the antiferromagnetic super exchange interaction. Please note: Since the thin NiO films are grown in (100) direction, the $\{$111$\}$ plane is not parallel to the sample surface. From the known in-plane lattice parameter $a$ and out-of-plane lattice parameters $c$, the spacing of the $\{$111$\}$ planes for a tetragonally distorted unit cell can be calculated by 
\begin{equation}
d_{hkl} = \left\{\frac{1}{a^2}\left(h^2+k^2\right)+\frac{1}{c^2}l^2\right\}^{-\frac{1}{2}},
\label{spacing1}
\end{equation} 

with $h = k = l = 1$. 

As can be clearly seen in Fig.~\ref{TN_and_2M_shift}(a), both quantities describing the magnetic characteristics ($\Delta\Omega_\text{2M}^0$ and $\Delta T_\text{N}$) exhibit a nearly linear dependence on $d_{111}$. Since $\Omega_\text{2M}^0$ and $T_\text{N}$ are both proportional to the antiferromagnetic exchange constant $J$, the observed monotonic relationships with the distance between the antiferromagnetically coupled $\{$111$\}$ lattice planes constitutes the expected behavior:\citep{massey1990a}
\begin{equation}
\frac{\hslash\Omega_\text{2M}^0}{k_{B}T_\text{N}} = \frac{\alpha}{\gamma} \\\
\quad\mathrm{with}\quad 
\label{Exchange1}
\end{equation} 
\begin{equation}
\hslash\Omega_\text{2M}^0 = \alpha J
\quad\mathrm{and}\quad 
k_{B}T_\text{N} = \gamma J,
\label{Exchange2}
\end{equation} 
where $\alpha$ and $\gamma$ are proportionality constants.

For thin epitaxial films the crucial question is whether or not the deviation from bulk NiO is caused by the lattice strain alone. As a matter of fact, the comparison to results obtained for bulk NiO under hydrostatic pressure \citep{massey1990a} reveals that the magnitude of the observed shifts shown in Fig.~\ref{TN_and_2M_shift}(a) cannot be explained only by the biaxial strain in the films [see shift of the N{\'e}el temperature expected according to Ref. \onlinecite{massey1990a} shown in Fig.~\ref{TN_and_2M_shift}(a) for comparison]. Furthermore, as shown in Fig.~\ref{TN_and_2M_shift}(b), the ratio $\hslash\Omega_\text{2M}^0\,/\,k_{B}T_\text{N}$ does not maintain the expected constant value $\alpha/\gamma$ (cf. Eq.~\ref{Exchange1}) for a super exchange interaction that depends only on the lattice strain\citep{massey1990a}, as indicated by the dotted line. This finding indicates that $\Omega_\text{2M}^0$ and $T_\text{N}$, i.e., the proportionality constants $\alpha$ and $\gamma$ (cf. Eq.~\ref{Exchange2}), depend in different manners on the growth conditions.

In order to explain the observed variation of the magnetic characteristics, we recall that the magnitude of $1-Q$ of the thin NiO films decreases with increasing in-plane strain (cf. Fig.~\ref{Q_and_2P_shift}) and, therefore, also the distance between $\{$111$\}$ planes. Consequently, we conclude that also the increasing lattice disorder contributes to the decrease in the antiferromagnetic exchange constant $J$. As already mentioned above, the disorder in the thin NiO films on MgO substrates is most likely dominanted by point defects which, in general, are expected to have a strong impact on the antiferromagnetic exchange between Ni$^{+2}$ ions mediated by oxygen orbitals.\citep{hutchings1972a} Consequently, the antiferromagnetic coupling becomes increasingly disturbed by a growing concentration of point defects. Our results indicate that the impact on the low-temperature magnon frequency is different from that on the N{\'e}el temperature (see Eq.~\ref{Exchange1} and Fig.~\ref{TN_and_2M_shift}). Further work is nescessary to explain this behavior.
\section{SUMMARY AND CONCLUSIONS}

In summary, Raman spectroscopy is demonstrated as a method to evaluate the magnetic characteristics of thin NiO films including a reliable approach for the determination of the N{\'e}el temperature. The influence of the growth conditions on the low-temperature magnon frequency and the N{\'e}el temperature is explained by the variation of the spacing between $\{$111$\}$ lattice planes due to biaxial lattice strain and a significant contribution of lattice disorder introduced by point defects, induced by low growth temperatures. Our results show that besides strain adjustment, e.g. by the proper choice of substrates, also the growth temperature can be used to engineer $T_\text{N}$.

\begin{acknowledgments}
We thank R. Goldhahn and M. Feneberg for providing the bulk NiO reference sample as well as Alberto Hernández-Mínguez for critically reading the manuscript. This work was performed in the framework of GraFOx, a Leibniz-ScienceCampus partially funded by the Leibniz association. J.\ F.\ and M.\ B.\ gratfully acknowledge the financial support by the Leibniz association.
\end{acknowledgments}


%

\end{document}